\documentclass[epj]{svjour}

\usepackage{graphicx}

\begin{document}

\newcommand{\icm}{\ensuremath{\mbox{cm}^{-1}}}
\newcommand{\pf}{$\rm (TMTSF)_2PF_6$}
\newcommand{\fso}{$\rm (TMTSF)_2FSO_3$}
\authorrunning{A. Pashkin et al.}
\titlerunning{Metal-insulator transition in (TMTSF)$_2$FSO$_3$ probed by infrared spectroscopy}
\sloppy

\title{Metal-insulator transition in the low-dimensional organic conductor (TMTSF)$_2$FSO$_3$ probed by infrared microspectroscopy}
\author{A. Pashkin\inst{1}\thanks{email: oleksiy.pashkin@physik.uni-augsburg.de} \and K. Thirunavukkuarasu\inst{1} 
\and Y.-L. Mathis\inst{2} \and W. Kang\inst{3} \and C. A. Kuntscher\inst{1}\thanks{email: christine.kuntscher@physik.uni-augsburg.de}}

\institute{Experimentalphysik\ II, Universit\"at Augsburg, 86159
Augsburg, Germany \and Institute for Synchrotron Radiation,
Forschungszentrum Karlsruhe, P.O. Box 3640, 76021 Karlsruhe,
Germany \and Department of Physics, Ewha Womans University, Seoul
120-750, Korea}

\date{Received: \today}

\abstract{We present measurements of the infrared response of the
quasi-one-dimensional organic conductor (TMTSF)$_2$FSO$_3$ along
(\textbf{E}$\|a$) and perpendicular (\textbf{E}$\|b'$) to the
stacking axis as a function of temperature. Above the
metal-insulator transition related to the anion ordering the
optical conductivity spectra show a Drude-like response. Below the
transition an energy gap of about 1500~\icm\ (185 meV) opens,
leading to the corresponding charge transfer band in the optical
conductivity spectra. The analysis of the infrared-active
vibrations gives evidence for the long-range crystal structure
modulation below the transition temperature and for the
short-range order fluctuations of the lattice modulation above the
transition temperature. Also we report about a new infrared mode
at around 710~\icm\ with a peculiar temperature behavior, which
has so far not been observed in any other (TMTSF)$_2X$ salt
showing a metal-insulator transition. A qualitative model based on
the coupling between the TMTSF molecule vibration and the
reorientation of electrical dipole moment of the FSO$_3$ anion is
proposed, in order to explain the anomalous behavior of the new
mode.
}
\maketitle

\section{Introduction}

The organic Bechgaard salts (TMTSF)$_2X$ consist of stacks of
planar TMTSF (tetramethyltetraselenafulvalene) molecules separated
by anions ($X$ = PF$_6$, AsF$_6$, ClO$_4$, Br, etc.). The charge
transport in these systems is restricted to the direction along the
molecular stacks, making the Bechgaard salts prime examples of
one-dimensional metals. However, on cooling down most of them
undergo a metal-insulator transition which prevents the onset of a
superconducting state \cite{Ishiguro98}. In Bechgaard salts with
noncentrosymmetric anions such as ReO$_4$, BF$_4$ or FSO$_3$ the
metal-insulator transition is related to the anion ordering
\cite{Pouget96}. It was furthermore demonstrated that in some cases the
metal-insulator transition can be suppressed by the application of
external pressure, leading to a superconducting ground state
\cite{Jerome04}.

The case of the anions $X$=FSO$_3$ in this class of materials is particularly
interesting, since these anions are noncentrosymmetric and in
addition possess a permanent electrical dipole moment. The first
study of the basic properties of \fso\ has been reported by Wudl et al.
in 1982 \cite{Wudl82}. Further studies have shown that this
compound has the highest superconducting transition temperature
(2.5 K at 8.5 kbar) among the Bechgaard salts. It was
proposed that this is due to the interaction of the
conducting electrons with the FSO$_3$ anion dipoles \cite{Lacoe83}. A
recent detailed study \cite{Jo03} revealed a very rich
pressure-temperature phase diagram of \fso\ with a variety of different
phases, which have not been completely identified up to now. Furthermore,
by magnetoresistance measurements a two-dimensional electronic
behavior was found in \fso\ under a pressure of around 6.2 kbar
\cite{Kang03}.

The interaction of the FSO$_3$ anions with each other via
long-range Coulomb forces and with the centrosymmetric surrounding
formed by the TMTSF cations tends to order the anions below a
certain temperature. The first-order structural phase transition
related to this anion ordering occurs at around $T_{MI}$=89 K in \fso\ at ambient
pressure. The change of the crystal structure modifies the electronic
band structure: The effective half-filled conducting band
splits into one filled and one empty band separated by an energy
gap, leading to a sharp metal-insulator transition
\cite{Lacoe83}. The structural analysis suggested a modulation of
the crystal structure with wavevector \textbf{q} = (1/2, 1/2, 1/2) below
the phase transition, which implies an antiferroelectric state
\cite{Moret83,Pouget96}. The ordering of the FSO$_3$ anions
modulates the lattice resulting in a new unit cell of size
$2a\times 2b\times 2c$. Thus, there are eight formula units of
\fso\ per unit cell in the low temperature phase. Correspondingly,
one can expect a splitting of each vibrational mode into up to eight
components \cite{Homes89}.

The ratio of the energy gap to the transition temperature in \fso\
is $\sim 12.5$ \cite{Wudl82}, which is appreciably higher than the
value 3.5 predicted by the mean-field theory for the Peierls
transition. Therefore, the metal-insulator in the Bechgaard salts
with non-centrosymmetric anions was attributed to a special type
of Peierls instability which originates from the anion-electron
coupling \cite{Jacobsen82}.

In this work we present the results of a temperature-dependent
polarized infrared reflectivity study of \fso\ single crystals in the far- and
mid-infrared frequency range, in order to characterize the change
of electronic and vibrational properties during the
metal-insulator transition at $T_{MI}$=89 K. This is the first
infrared spectroscopic investigation of the compound \fso. 
Our results allow a direct determination of the charge gap in
the insulating state. Furthermore, we determined and analyzed the
behavior of the vibrational modes during the metal-insulator
transition, which can clarify details of the dipolar ordering.

\section{Experimental}

\fso\ single crystals were grown by standard electrochemical
techniques from TMTSF molecules and tetrabutylammonium-FSO$_3$.
The studied samples have a needle-like shape, with a size of
approximately $2\times 0.2\times 0.1$~mm$^3$. The samples were
mounted on a cold-finger CryoVac Konti-Mikro cryostat. The actual
measuring temperature was controlled by a sensor attached in
direct vicinity of the sample. The measurements were performed at
the infrared beamline of the synchrotron radiation source ANKA.
The polarized infrared reflectivity was measured in the range 150
- 10000 \icm\ using a Bruker IRscope II microscope attached to a
IFS66v/S spectrometer. The frequency resolution was 1~\icm\ for
all measured spectra. Optically transparent TPX and KBr cryostat
windows were used for the measurements in the far- and
mid-infrared frequency range, respectively.
\begin{figure}
  \includegraphics[width=0.95\columnwidth]{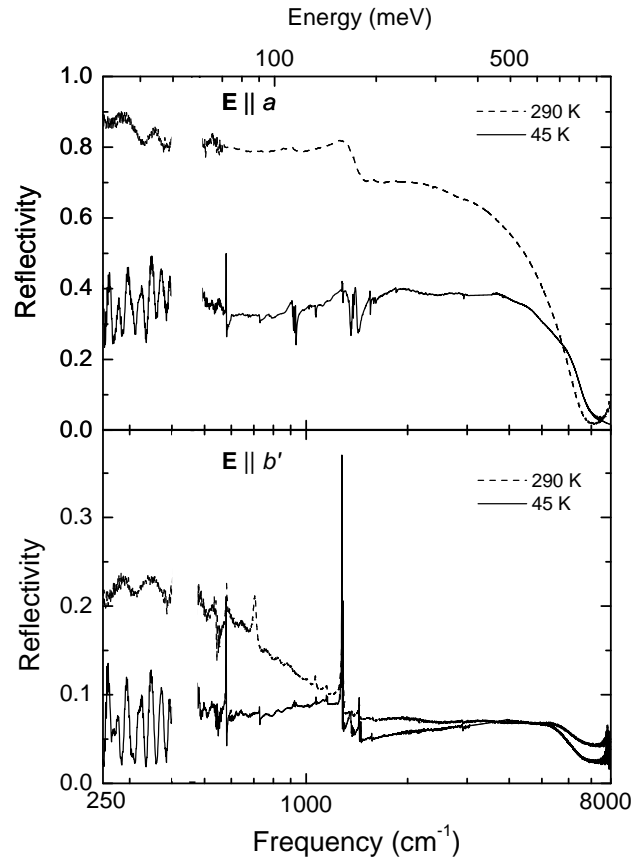}\\
  \caption{Reflectivity spectra of \fso\ above and below the metal-insulator
  transition for \textbf{E}$\| a$ and \textbf{E}$\| b'$.
  \label{fig:refl}}
\end{figure}

\section{Results and discussion}

\subsection{Electronic properties}

The reflectivity spectra of \fso\ above and below the
metal-insulator transition temperature of 89~K for both
polarizations \textbf{E}$\| a$ and \textbf{E}$\| b'$ (along and
perpendicular to the stacking axis, respectively) are shown in
Fig.~\ref{fig:refl}. The reflectivity data in the spectral region
at around 450~\icm\ are affected by the absorption features of the
far-infrared TPX cryostat window and are therefore not shown.

At 290~K the reflectivity of the sample along the stacking axis
\textbf{E}$\| a$ demonstrates a typical Drude behavior (growth up
to 1 when frequency tends to zero). In contrast, at 45~K , i.e.,
below $T_{MI}$, the reflectivity is almost frequency independent
below 1000~\icm, which is typical for an insulating state.

The interference fringes observed below 400~\icm\ in the spectra
for both polarizations are due to the partial transparency of the
sample in the insulating phase. Perpendicular to the stacking axis
(\textbf{E}$\| b'$), the optical reflectivity and conductivity is
much lower than along the $a$ axis. Nevertheless, the observed
changes during the metal-insulator transition are similar to 
those of the \textbf{E}$\| a$
direction. These results demonstrate the opening of an energy gap
at the Fermi level for both studied directions.

\begin{figure}
  \includegraphics[width=1\columnwidth]{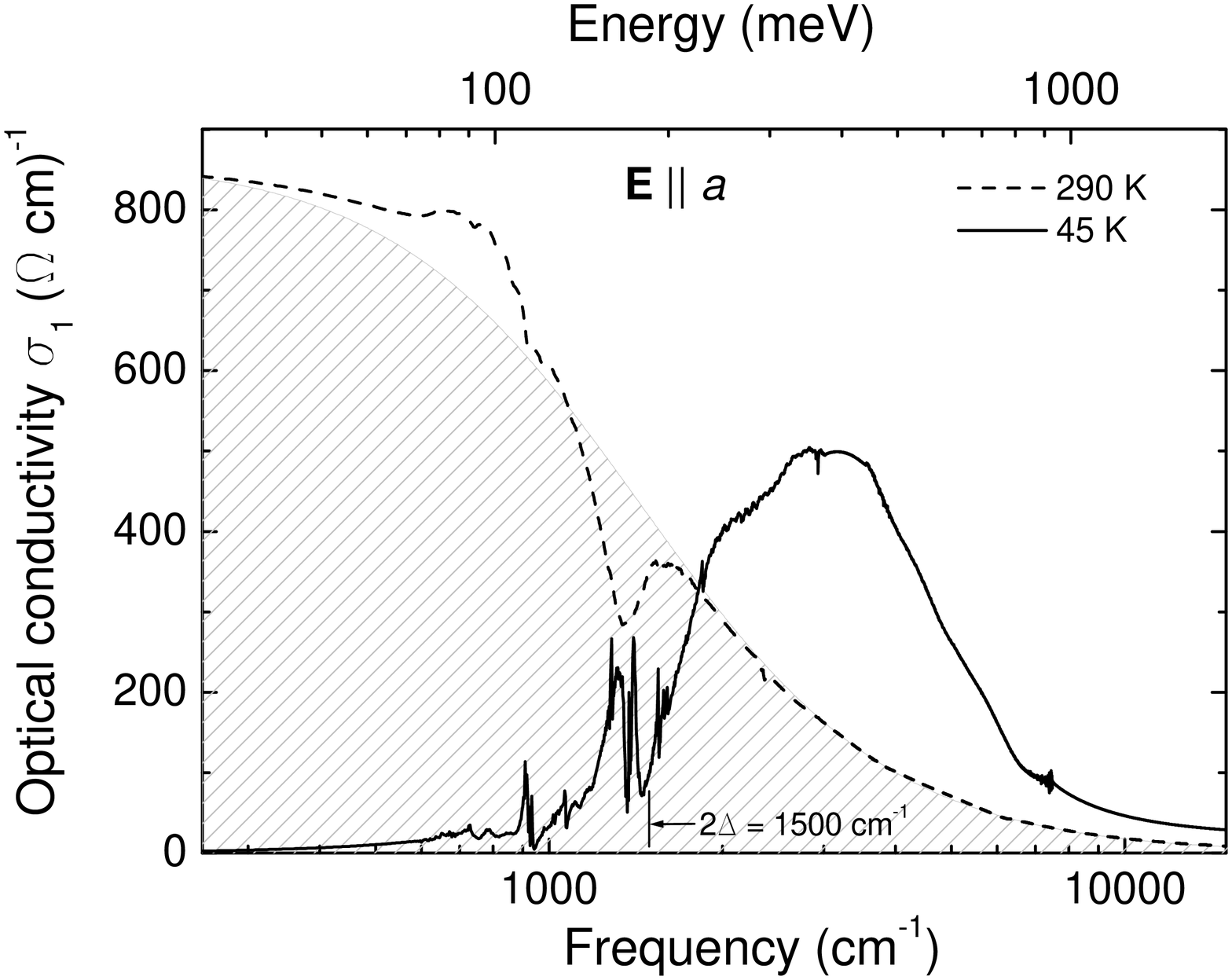}\\
  \caption{\textbf{E}$\| a$ optical conductivity spectra of \fso\ above and below the metal-insulator transition
  at $T_{MI}$= 89 K. Hatched area depicts the Drude model fit of the
  high temperature optical conductivity.
  \label{fig:sigma}}
\end{figure}

The dramatic effect of the temperature decrease on the electronic
properties of \fso\ are more directly seen in the optical
conductivity spectra. The \textbf{E}$\| a$ optical conductivity
$\sigma_1(\omega)$ of \fso\ in the insulating (at 45~K) and
conducting phase (at 290~K) obtained by means of Kramers-Kronig
analysis is shown in Fig.~\ref{fig:sigma}. The dominating feature
of the spectrum at 45~K is a strong charge transfer band due to
electronic transitions across the gap. The arrow shows the band
gap (1500~\icm) obtained from the published temperature-dependent
dc resistivity measurements \cite{Wudl82}. Obviously, the
agreement of this value with the onset of the optical interband
transition is very good. On the other hand, the optical
conductivity at room temperature is mostly dominated by the Drude
response of the free carriers. The corresponding fit using the
Drude model is shown as the hatched area in Fig.~\ref{fig:sigma}.
Obviously, the Drude model provides a good description of the
measured room-temperature spectrum excluding the 
electron-molecular vibration (emv) antiresonance
modes. The plasma frequency $\omega_p=8660$~cm$^{-1}$ and the
scattering rate $\rm \Gamma\simeq 1450$~cm$^{-1}$ obtained from
the fit agree well with the Drude model parameters reported for
other TMTSF salts \cite{Jacobsen83}. The obtained value of the dc
conductivity, $\sigma_{dc}\simeq 860$~$\rm
(\Omega cm)^{-1}$, is in reasonable agreement with the dc and
microwave conductivity values of 1600 and 300~$\rm (\Omega
cm)^{-1}$, respectively, reported by Wudl et al.~\cite{Wudl82}.

\subsection{Vibrational modes}

The TMTSF molecule with the point group symmetry $D_{2h}$ has in
total 72 local vibrational modes classified according to the
following representations \cite{Meneghetti84}
\begin{eqnarray}
 \Gamma_{D_{2h}}=(12a_g+11b_{3g}+11b_{1u}+11b_{2u})\nonumber\\
  +(6b_{1g}+7b_{2g}+7a_u+7b_{3u}),
  \label{eq:TMTSF}
\end{eqnarray}
where the vibrations in the first brackets are polarized in the
molecular plane (perpendicular to the stacking $a$ axis) and the
vibrations in the second brackets are polarized out of the plane
(along the stacking axis $a$). The symmetric (gerade) vibrations
are Raman active and the asymmetric (ungerade) vibrations are
infrared active excluding the $a_u$ silent modes. Some of the
totally symmetric $a_g$ Raman modes are expected to appear in the
infrared spectra for \textbf{E}$\|a$ due to efficient
emv coupling in the modulated stacking structure \cite{Jacobsen83,Homes90}.

The tetrahedral FSO$_3$ anion has $C_{3v}$ point group symmetry
which gives in total nine vibrational modes
\begin{equation}
  \Gamma_{C_{3v}}=3a_1(z,x^2+y^2,z^2)+3e(x,y,x^2-y^2,xy,yz,xz),
  \label{eq:FSO3}
\end{equation}
where $e$ species correspond to the doublets. Thus, in the
infrared spectra one expects six modes, with the $3a_1$ and $3e$
modes being polarized along and perpendicular to the polar axis of
the anion, respectively.

\begin{table}
\caption{\label{tab:modes-a} The eigenfrequencies and assignment
of some vibrational modes observed in \fso\ for \textbf{E}$\| a$
at 45~K below $T_{MI}$. All numbers are in \icm.}
\begin{tabular}{ccc}
\hline\noalign{\smallskip}
45 K & calculated frequency\footnote{for
TMTSF$^{0.5+}$ according
to \cite{Meneghetti84,Homes90}, for FSO$_3$ according to \cite{Nakamoto86b}} & assignment\\
\noalign{\smallskip}\hline\noalign{\smallskip}
580 & 571 & $\nu_3(a_1)$ FSO$_3$ \\
\\
728 & 702 & $\nu_{51}(b_{2u})$ \\
\\
902, 911, 915,  & 916 & $\nu_8(a_g)$ \\
917, 924, 932 & & \\
\\
1020, 1031, 1036 & 1060 & $\nu_7(a_g)$\\
\\
1067, 1072 & 1060 & $\nu_7(a_g)$ \\
\\
1362, 1450\footnote{strong antiresonance modes} & 1469 & $\nu_4(a_g)$ \\
\\
1354, 1364, 1369 & 1369 & $\nu_6(a_g)$ \\
1373, 1379, 1385 &  &  \\
\\
1550, 1584, 1606 & 1596 & $\nu_3(a_g)$ \\
\\
1847, 1854, 1863 & 1863 & $\nu_3(a_g)+\nu_{11}(a_g)$ \\
\noalign{\smallskip}\hline
\end{tabular}
\end{table}

In this section we want to concentrate on the changes in the
infrared phonon spectra for both polarizations across the metal-insulator
transition. For \textbf{E}$\|a$ several $a_g$ vibrations of the
TMTSF molecules become infrared active in the insulating phase.
This is due to the effective emv coupling of these
vibrations to the on-chain charge transfer band in the structure
modulated due to the anion ordering. The list of the new modes
observed below the transition together with their tentative
assignment is given in Table~\ref{tab:modes-a}. Most of them are
emv coupled $a_g$ modes polarized in the molecular plane or their
combination as a triplet at around 1850~\icm. The $\nu_4(a_g)$
mode involving the central C=C bond stretching is known to have
especially strong emv coupling and therefore it appears as a
strong antiresonance mode in the optical conductivity spectrum. It should
be pointed out that the observed appearance of $a_g$ modes for
\textbf{E}$\|a$ in the ordered phase is typical only for
(TMTSF)$_2X$ compounds with non-centrosymmetric anions. In
comparison, (TMTTF)$_2X$ salts possess a stronger stack
dimerization, resulting in the emv coupling of the $a_g$ modes
already in the disordered phase, and therefore the anion ordering transition
causes only a frequency shift and an intensity change of the emv
coupled modes \cite{Garrigou84}.

\begin{figure}[t]
  \includegraphics[width=1\columnwidth]{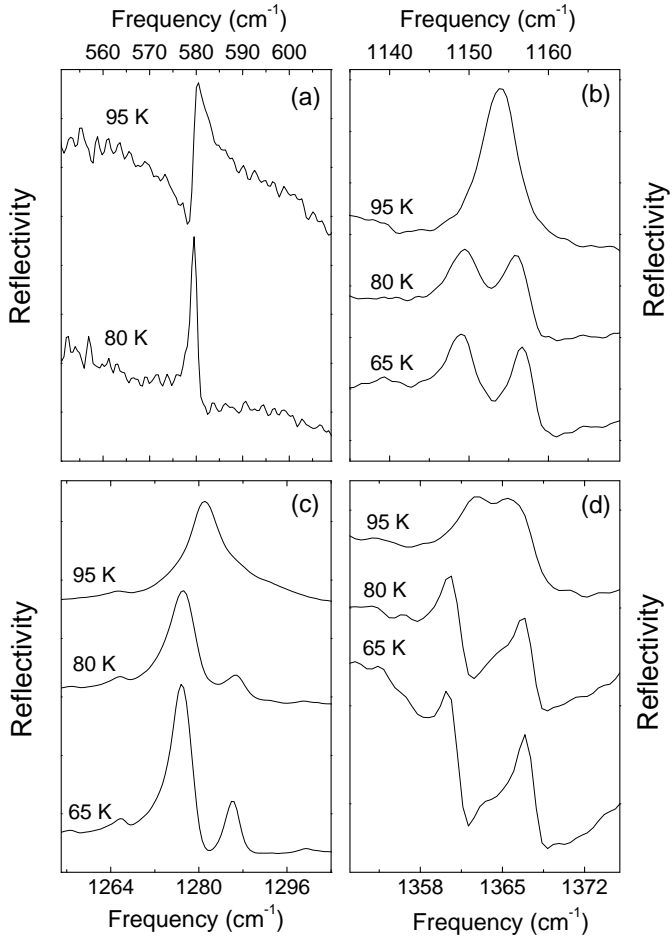}\\
  \caption{Reflectivity spectra (shifted for clarity) of some phonons which experience 
  changes during the metal-insulator transition at 89~K:
   (a) vibration polarized along the $a$ axis; (b)-(d) vibrations polarized along the $b'$ axis.
    \label{fig:modes}}
\end{figure}

The $\nu_3(a_1)$ vibrational mode of the FSO$_3$ anion at 580~\icm\
is observed for the whole studied temperature range. However, the
lineshape of this mode in the metallic phase above $T_{MI}$ is
inverted with respect to the insulating phase [see
Fig.~\ref{fig:modes}(a)], since the background dielectric constant is
negative as expected for highly conducting metals at low
frequencies. Such a change is a clear evidence for the suppression
of the Drude conductivity in the insulating phase of \fso.

\begin{figure}
\begin{center}
  \includegraphics[width=0.9\columnwidth]{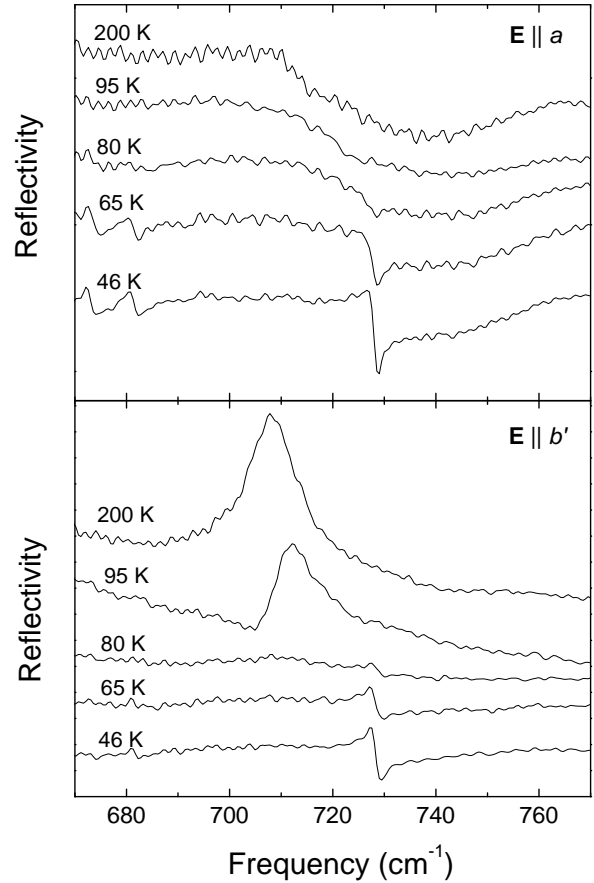}\\
  \caption{Reflectivity spectra (shifted for clarity) of the vibration at around 
  710~\icm\ at different temperatures for \textbf{E}$||a$ and \textbf{E}$||b'$.
    \label{fig:mode710}}
    \end{center}
\end{figure}

The mode at 728~\icm\ observed for temperatures below
$T_{MI}$ is particularly interesting, since its intensity
gradually increases on temperature decrease (see
Fig.~\ref{fig:mode710}). A similar behavior is found for
the polarization perpendicular to the stacks, \textbf{E}$\|b'$.
Moreover, above the transition temperature a strong asymmetric
mode is seen at 710~\icm. This mode shifts to lower frequencies
and gets stronger with increasing temperature. This mode has not
been observed in any other earlier study of the Bechgaard salts.
Therefore, it would be natural to assign it to a vibration of
FSO$_3$ anion. However, such an assignment would be in
contradiction to the experimental observations, since: (i) the
$\nu_5(e)$ and $\nu_2(a_1)$ vibrations of FSO$_3$ located close to the
observed mode have frequencies which are by more than 100~\icm\
higher or lower \cite{Nakamoto86b}; (ii) the intensity of the
anion vibration should not vanish at the order-disorder transition
point. Thus, one has to attribute the modes at around 710 and
728~\icm\ to vibrations of the TMTSF molecules. We suggest that
both modes originate from the $\nu_{51}(b_{2u})$ in-plane
vibration of the TMTSF molecule. According to the
normal-coordinate analysis \cite{Meneghetti84,Eldridge91} its
frequency for a free TMTSF$^{0.5+}$ cation is 702~\icm.

\begin{figure}
  \includegraphics[width=1\columnwidth]{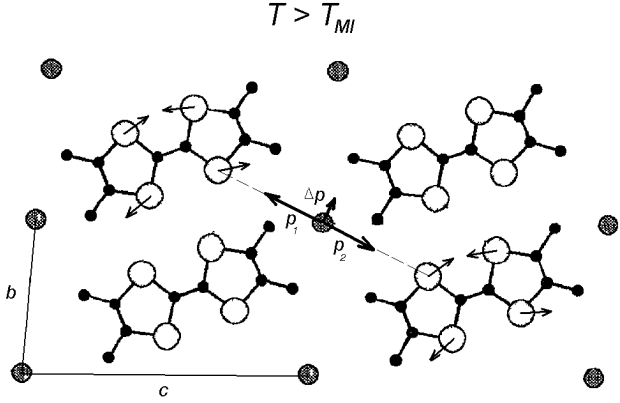}\\
  \caption{Schematic illustration of the $\nu_{51}(b_{2u})$ vibration coupled 
  to the reorientation of the FSO$_3$ electrical dipole
  moment. The projection of the crystal structure on the $b-c$ plane is shown.
  Only the Se (large open
  circles) and C (small filled circles) atoms of the TMTSF molecules are
  presented, together with the displacements of the Se atoms. 
  The grey filled circles between molecules denote the positions of
  the FSO$_3$ anions, the bold arrows show the two possible orientations
  of the anion dipole moment (\textit{\textbf{p}}$_1$ and \textit{\textbf{p}}$_2$). 
  Because of the symmetry properties of the $b_{2u}$ vibration the 
  reorientation of the electrical dipole moment leads to a change of polarization 
  $\Delta$\textit{\textbf{p}} in the perpendicular direction for
  any orientation of the anion dipole moment.
    \label{fig:b2u}}
\end{figure}

The corresponding atomic movements involve stretching of the Se-C
side bond and rocking of the adjacent methyl group. For the
$b_{2u}$ vibration the inversion symmetry of the molecule is not
preserved, causing its infrared-activity for the polarization
perpendicular to the stacks. However, it is known that in
(TMTSF)$_2X$ salts the dipole moment corresponding to the
$\nu_{51}(b_{2u})$ vibration is very small, and therefore this
mode can hardly be detected even for \textbf{E}$\|b'$ where it
should have the strongest intensity \cite{Eldridge91}. Nevertheless,
in \fso\ this mode is particularly strong even at room
temperature. This finding can be explained by the electrical dipole
of the FSO$_3$ anion
pointing towards the Se-F bond. Similar to other
non-centrosymmetric anions (ReO$_4$, ClO$_4$ etc.) the FSO$_3$
anion has two possible symmetrically equivalent orientations for which
the dipole moment points towards the Se atoms of the neighboring
TMTSF molecules. This situation is sketched in Fig.~\ref{fig:b2u},
where \textit{\textbf{p}}$_1$ and \textit{\textbf{p}}$_2$ are two
possible orientations of the FSO$_3$ electrical dipole moment.
During the vibration the dipole moment of the anion follows the
position of the Se atom. Due to the symmetry properties of the
$b_{2u}$ vibration the nearest Se atoms on both sides of the anion
move in the same direction. Thus, for both possible orientations
of the dipole the $b_{2u}$ vibration results in a change of the
average polarization along the direction of
$\Delta$\textit{\textbf{p}} (Fig.~\ref{fig:b2u}).

The described coupling mechanism between the $b_{2u}$ vibration
and the dipole moment of the anion in \fso\ should lead to a
strong enhancement of the infrared strength of the
$\nu_{51}(b_{2u})$ vibration for \textbf{E}$\|b'$, since
$\Delta$\textit{\textbf{p}} has the largest projection along this
direction. On the other hand, $\Delta$\textit{\textbf{p}} is
perpendicular to the stacking axis and the $b_{2u}$ mode should
not appear for \textbf{E}$\|a$. This is indeed observed in our
experiment above the transition temperature. Below the transition
the long-range order of the anion sublattice builds up. Then the
anion dipole moment orientation is determined by the modulation of
the whole lattice and it is not dependent on the movement of
neighboring TMTSF molecules, i.e., the $\nu_{51}(b_{2u})$
vibration is decoupled from the FSO$_3$ anions. Therefore, its
intensity should drop abruptly below $T_{MI}$, in agreement with
our observations (see Fig.~\ref{fig:mode710}). Moreover, the
observed decrease of the intensity of the coupled $b_{2u}$ mode at
around 710~\icm\ at 95~K compared to higher temperatures can be
explained by taking into account short-range order fluctuations above
the transition, evidence for which is also given below. Indeed, in
the large enough dynamical regions where the anions are ordered,
the coupling is suppressed and therefore the strength of the
$\nu_{51}(b_{2u})$ should decrease.

On cooling down below $T_{MI}$ a vibration appears again at
somewhat higher frequency (728~\icm) for \textbf{E}$\|b'$ and its strength gradually
increases with decreasing temperature. We suggest that this is the
same $\nu_{51}(b_{2u})$ vibration described above. Since it is
decoupled from the anion sublattice, its frequency is expected to
increase abruptly below the transition. The increase in strength
for both polarizations should be obviously related to the
temperature dependence of the order parameter (i.e., the degree of
lattice modulation). One of the possible mechanisms can be the emv
coupling of the $\nu_{51}(b_{2u})$ vibration to the charge
transfer bands along $a$ and $b'$ directions. However, the
detailed picture of this emv coupling is not clear, since the
symmetry of $b_{2u}$ mode does not allow such kind of coupling.
One can speculate that the electric field of the FSO$_3$ dipoles
in the ordered phase distorts the TMTSF molecules making them
non-centrosymmetric. Then the emv coupling may become allowed for
the $b_{2u}(\nu_{51})$ mode.

\begin{table}
\caption{\label{tab:modes-b} The eigenfrequency, width (given in bracket), 
and assignment of some vibrational modes observed for \textbf{E}$\|
b'$ at selected temperatures. All numbers are in \icm.}
\begin{tabular}{cccc}
\hline\noalign{\smallskip}
95 K & 80 K & 45 K & assignment\\
\noalign{\smallskip}\hline\noalign{\smallskip}
580 (1.3) & 580 (0.9) & 580 (0.8) & $\nu_3(a_1)$ FSO$_3$ \\
\\
710 (7.1) & 728 & 728 (2.0) & $\nu_{51}(b_{2u})$ \\
\\
1154 (5.0) & 1150 (3.4) & 1150 (3.3) & $\nu_{48}(b_{2u})$ \\
 & 1157 (3.8) & 1158 (2.5) &  \\
\\
1280 (5.1) & 1276 (4.2) & 1276 (2.1) & $\nu_{4}(e)$ FSO$_3$\\
1288 (16) & 1286 (4.4) & 1286 (1.7) &  \\
\\
1363 (3.2) & 1361 (2.8) & 1361 (1.3) & $\nu_{47}(b_{2u})$ \\
1366 (4.3) & 1367 (2.4) & 1367 (1.8) &  \\
\noalign{\smallskip}\hline
\end{tabular}
\end{table}

Noticeable changes in the phonon mode spectra across the
metal-insulator transition are observed for \textbf{E}$\|b'$. The
list of the parameters of these modes at temperatures above and
below $T_{MI}$ is given in Table~\ref{tab:modes-b}. An obvious
splitting into two components is seen for the $\nu_{48}(b_{2u})$
mode at 1154~\icm\ [see Figure~\ref{fig:modes}(b)]. In addition, the
damping of the split components directly below $T_{MI}$ is lower
than the damping of the single component directly above the
transition (see Table~\ref{tab:modes-b}). This difference is
probably related to the precursor short-range order fluctuations
above the transition, which can induce a small splitting already
in the disordered phase. An evidence for such fluctuations was
found in x-ray diffuse scattering experiments
\cite{Moret83,Pouget96}. This effect is even more clearly seen in
the splitting of two other modes: the doublet $\nu_{4}(e)$
vibration of the FSO$_3$ anion at 1280~\icm\
[Fig.~\ref{fig:modes}(c)] and the $\nu_{47}(b_{2u})$ mode at around
1365~\icm\ [Fig.~\ref{fig:modes}(d)]. For each of these modes above
$T_{MI}$ one can resolve two weakly split components. However,
below $T_{MI}$ the splitting abruptly increases and the damping
decreases (see Table~\ref{tab:modes-b}) indicating the onset of
long-range order. Since the described effect is observed not only
for the FSO$_3$ anion vibration but also for two vibrations of the
TMTSF cation, we can conclude that the short-range order
fluctuations involve the modulation of the whole \fso\ lattice and
not only the anion sublattice.

\section{Conclusion}
We have performed an infrared spectroscopic study of the
metal-insulator transition in \fso. The obtained optical
conductivity spectra for \textbf{E}$\|a$ show a Drude-like
conductivity above the anion ordering temperature and a charge
transfer band formed below the transition. The onset of this band
is in agreement with the energy gap value of 1500~\icm\ obtained
from transport measurements \cite{Wudl82}.

The analysis of the infrared-active vibrations leads to the
following conclusions: (i) the crystal structure modulation below
the metal-insulator transition leads to a strong emv coupling of several
$a_g$ vibrations which therefore become infrared-active; (ii)
short-range order fluctuations of the FSO$_3$ anions and the
corresponding lattice modulation exist above the transition
temperature, as it is seen from the splitting of some
infrared-active modes for \textbf{E}$\|b'$; (iii) a new infrared-active
mode located at around 710~\icm\ with a peculiar temperature behavior is
detected and assigned to the coupling between the $b_{2u}$ TMTSF
molecule vibration and the electrical dipole moment of the
FSO$_3$ anion. The latter feature has not been observed in any other
(TMTSF)$_2X$ salt showing a metal-insulator transition. This
points out the important role of the electrical dipole moment of
the anion on the structural and dynamical properties of the \fso\
salt.

\section{Acknowledgements}
We acknowledge the ANKA Angstr\"omquelle Karlsruhe for the provision
of beamtime and thank M. S\"upfle, D. Moss, and B. Gasharova for technical
assistance at the ANKA IR beamline. The financial support of the DFG
(Emmy Noether-program) is acknowledged.

\bibliographystyle{epj}
\bibliography{Bechgaard2}

\end{document}